\documentclass{INTERSPEECH2023}

\usepackage{csquotes}
\usepackage{xcolor}
\usepackage{soul}
\usepackage{subcaption}
\usepackage{multirow}

\interspeechcameraready 

\title{Towards Robust FastSpeech 2 by Modelling Residual Multimodality}
\name{Fabian Kögel, Bac Nguyen, Fabien Cardinaux}
\address{Sony Europe B.V., Stuttgart Laboratory 1, Germany}
\email{Fabian.Koegel@sony.com}

\begin{document}

\maketitle
\begin{abstract}
    State-of-the-art non-autoregressive text-to-speech (TTS) models based on FastSpeech 2 can efficiently synthesise high-fidelity and natural speech. 
    For expressive speech datasets however, we observe characteristic audio distortions. %
    We demonstrate that such artefacts are introduced to the vocoder reconstruction by over-smooth mel-spectrogram predictions, which are induced by the choice of mean-squared-error (MSE) loss for training the mel-spectrogram decoder.
    With MSE loss FastSpeech 2 is limited to learn conditional averages of the training distribution, which might not lie close to a natural sample if the distribution still appears multimodal after all conditioning signals. 
    To alleviate this problem, we introduce TVC-GMM, a mixture model of Trivariate-Chain Gaussian distributions, to model the residual multimodality.
    TVC-GMM reduces spectrogram smoothness and improves perceptual audio quality in particular for expressive datasets as shown by both objective and subjective evaluation.
\end{abstract}

\section{Introduction}

The distribution of natural speech signals is multimodal since the same content can be spoken in many different ways and dependent since it is continuous in time and coherent within a speaking voice~\cite{zenDeepMixtureDensity2014}.
Tractably modelling this complex distribution is a fundamental challenge in statistical parametric speech synthesis (SPSS)~\cite{zenStatisticalParametricSpeech2013}.
A common problem are audio artefacts due to over-smooth predictions caused by naive modelling assumptions~\cite{henterMeasuringPerceptualEffects2014}.
Ultimately, SPSS balances approximations of the natural speech distribution with capabilities of available modelling methods and computing resources.
This has led to various families of TTS models that achieve high-quality speech synthesis through autoregressive factorization~\cite{oordWaveNetGenerativeModel2016, liNeuralSpeechSynthesis2019}, end to end (E2E) training~\cite{wangTacotronEndtoEndSpeech2017}, generative adversarial loss~\cite{donahueEndtoEndAdversarialTexttoSpeech2021, nguyenDifferentiableDurationModeling2023}, variational autoencoding~\cite{kimConditionalVariationalAutoencoder2021} or normalizing flows~\cite{kimGlowTTSGenerativeFlow2020}.

In the family of non-autoregressive architectures, FastSpeech 2~\cite{renFastSpeechFastRobust2019, renFastSpeechFastHighQuality2021} is widely popular for its high-quality and fast parallel generation, small data requirements and inherent controllability. 
Its two stage pipeline separates the acoustic model from the vocoder and reduces training time by re-using pre-trained vocoders. %
Further, as discussed by~\cite{renRevisitingOverSmoothnessText2022}, its explicit modelling of pitch, energy and phoneme duration simplifies the distribution to be learned and reduces the over-smoothness and allows for manual control during generation.
However, characteristic audio artefacts can still be observed with models from this family - in particular with more expressive and multi-speaker datasets.
Although this problem has been treated by fine-tuning the vocoder or training it jointly with the acoustic model in an E2E architecture~\cite{donahueEndtoEndAdversarialTexttoSpeech2021, limJETSJointlyTraining2022, nguyenDifferentiableDurationModeling2023}, we believe this only masks the underlying issue by tuning to a particular dataset using high training efforts. %
We find that the conditioning is not sufficient for expressive datasets and there is residual multimodality that in conjunction with the inherent assumption of unimodality of the MSE loss function is the root cause for the audio degradation.
Adding more conditioning or fine-tuning for individual datasets is not sustainable for the application in controllable expressive speech synthesis with intended large diversity of prosody.

In this paper we propose a mixture model to model this residual multimodality as a compromise between the more powerful distribution mapping approaches (data- and computation-intensive and difficult to optimise approaches of GAN and normalizing flows) and the controllability, fast/data-efficient training and generation speed of the FastSpeech 2 architecture.

We make the following contributions:
(1) we demonstrate how smoothness causes artefacts across vocoder architectures for FastSpeech 2,
(2) we propose TVC-GMM\footnote{code at \href{}{https://github.com/sony/ai-research-code/tvc-gmm}}, a novel trivariate-chain gaussian mixture modelling layer which can tractably model the residual multimodality in the spectrogram distribution of non-autoregressive two-stage TTS models,
(3) we show that TVC-GMM reduces spectrogram smoothness and improves perceptual audio quality in objective and subjective evaluation in particular for expressive datasets, while maintaining fast and data-efficient training and generation speed.

\begin{figure}[t!]
    \centering
    \includegraphics[width=0.85\columnwidth,trim= 0 12 0 12,clip]{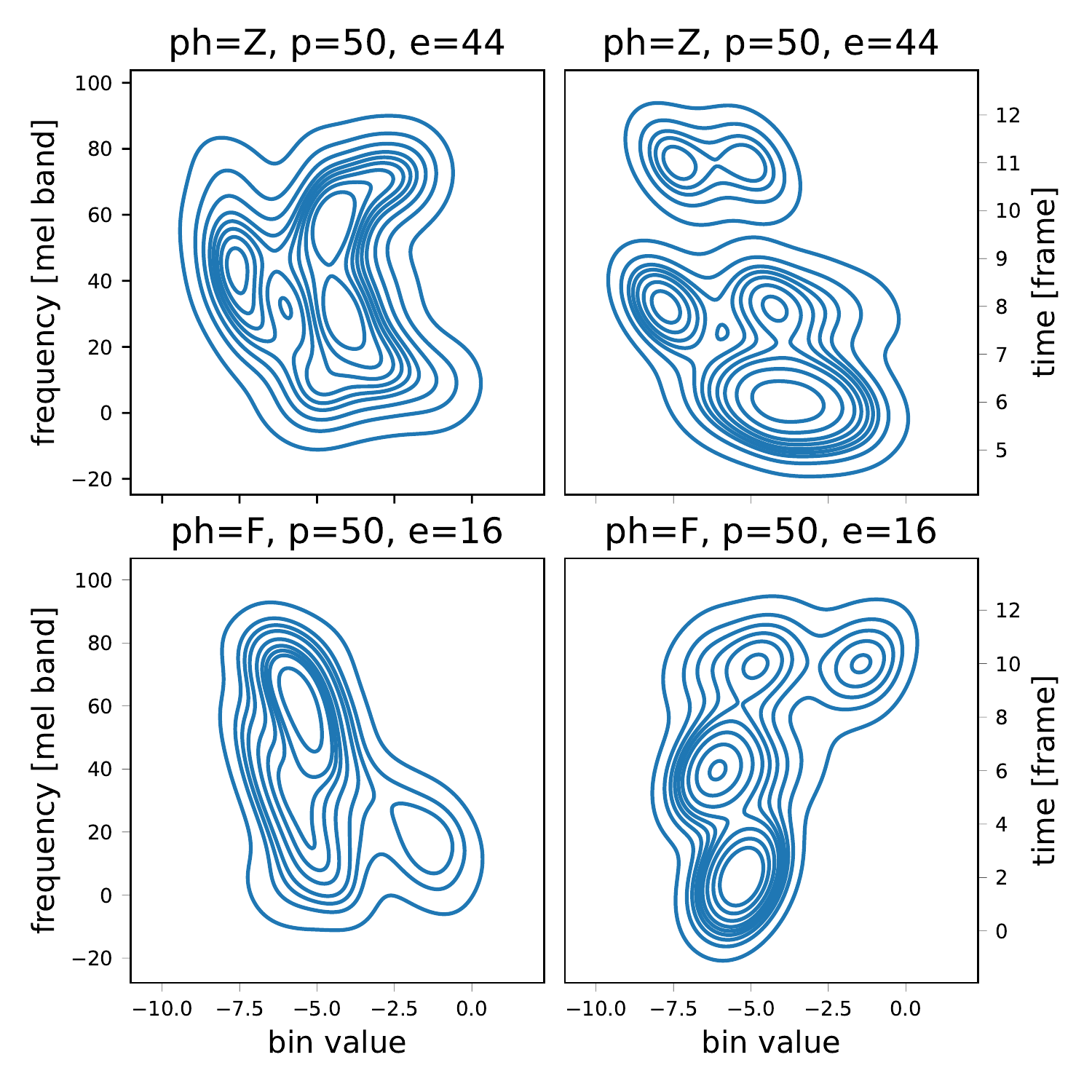}
  \caption{Examples of residual multimodality observed in marginal distributions over time and frequency for phoneme $ph$ after all pitch $p$ and energy $e$ conditioning in FastSpeech 2.}
  \label{fig:distributions}
\end{figure}

\section{Residual Multimodality}
\label{section:mm_dist}

We introduce the term residual multimodality for the gap in complexity between the actual distribution to be learned and the assumed distribution in the loss function.
Although deep neural models are universal approximators, the distribution that can practically be learned depends on the loss to be optimised~\cite{hornikMultilayerFeedforwardNetworks1989}.
FastSpeech 2 deploys MSE loss, which optimises training data likelihood only under an assumed gaussian distribution~\cite{bishopMixtureDensityNetworks1994} and thus relies on conditioning on external control signals to simplify the complex speech distribution.
In Figure \ref{fig:distributions}, we visualise the remaining distribution of our training data after all conditioning in FastSpeech 2 - that is speaker label, pitch bin and energy bin.
We show the estimated density of the spectrogram value distribution marginalized over time (left) or frequency (right) for selected phonemes $ph$ and pitch/energy bins $p$ and $e$ in the train split of the single-speaker LJSpeech dataset.
Even within single frequency bands, we observe multiple modes.
Additionally, we observe large variances in unimodal areas, which might indicate dependencies between bins.
The residual multimodality and dependencies that appear despite conditioning can cause over-smoothing artefacts as we show in Section \ref{sec:vocoder_degradation}.
Adding even more conditioning is unsustainable and distribution mapping approaches like normalizing flows add computation and data requirements, therefore we commence with modelling the residual multimodality.

\section{Trivariate-Chain Gaussian Mixture Modelling (TVC-GMM)}

\begin{figure}[th]
    \centering
    \includegraphics[width=\columnwidth]{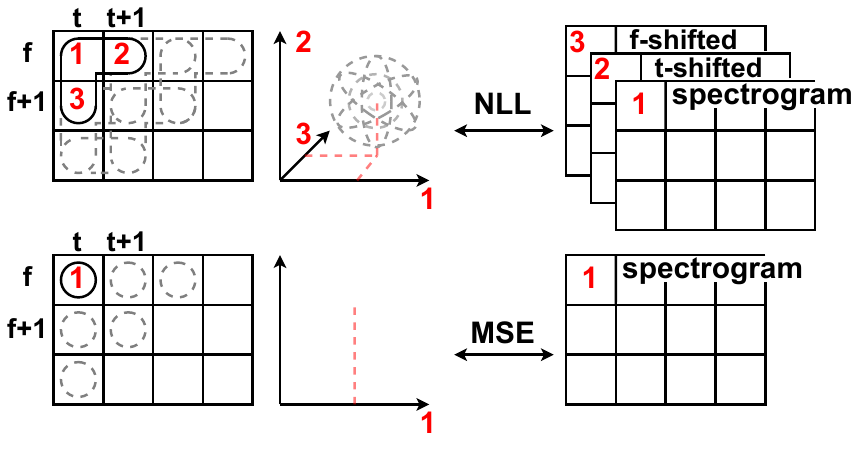}
    \caption{Illustration of approaches. TVC-GMM (top) models dependencies of adjacent spectrogram bins by a trivariate gaussian chain while FastSpeech 2 (bottom) only models means.}
    \label{fig:tvc_gmm}
\end{figure}

Let $\mathbf{Y} \in \mathbb{R}^{T \times F}$  be a random variable representing the mel-spectrograms, where $T$ and $F$ are the length and number of frequency bins.
Given some conditioning linguistic features $\mathbf{L}$ and control inputs $\mathbf{C}$, the goal for the acoustic model is to estimate the true distribution of mel-spectrograms by maximising the likelihood of the training data under the parametrized distribution $p_\mathbf{\theta}(\mathbf{Y}|\mathbf{L},\mathbf{C})$.
FastSpeech 2 achieves this by minimising the mean squared error (MSE) between the predicted and ground-truth spectrograms, which implicitly assumes the values of individual time/frequency bins $y^{t,f}$ to be the means of unimodal independent gaussian distributions~\cite{bishopMixtureDensityNetworks1994}.
As discussed in section \ref{section:mm_dist}, this assumption does not hold and learning the mean of the residual multimodal distribution leads to the over-smoothness problem in TTS.
To alleviate this problem, we propose TVC-GMM, a mixture model of Trivariate-Chain Gaussian distributions.

\subsection{Trivariate-Chain Gaussian distribution}
\label{sec:tvc_dist}

We start by addressing the dependence between adjacent bins. 
Ideally, we would like to model the full $T\times F$ dimensional spectrogram distribution, but this requires a prohibitively large covariance matrix and fixed $T$ and $F$.
Instead, we break it down and introduce a Trivariate-Chain Gaussian distribution. 

Let $\tilde{\mathbf{Y}} \in \mathbb{R}^{T\times F \times 3}$ be a random variable where each element $\tilde{\mathbf{Y}}^{t,f}$ is a triplet of neighbouring spectrogram bins $\mathbf{Y}^{t, f}$, $\mathbf{Y}^{t + 1, f}$, and $\mathbf{Y}^{t, f + 1}$ modelled by a trivariate Gaussian distribution, i.e. 
\begin{align}
    \tilde{\mathbf{Y}}^{t, f} \sim \mathcal{N}(\mathbf{\mu}^{t,f}, \mathbf{\Sigma}^{t,f}) \,,
\end{align}
where $\mathbf{\mu}^{t,f} \in \mathbb{R}^3$ and $\mathbf{\Sigma}^{t,f} \in \mathbb{R}^{3\times 3}$ denote the mean and covariance matrix. 
The $\tilde{\mathbf{Y}}^{t, f}$ are chained together by their overlapping targets and provide a way to represent local variances dependent on correlations between time and frequency steps.
Our objective is to learn the parameters $\mathbf{\mu}^{t,f}$ and $\mathbf{\Sigma}^{t,f}$ and to sample from $\tilde{\mathbf{Y}}$ instead of directly predicting individual spectrogram values as FastSpeech 2 does.

\subsection{Trivariate-Chain Gaussian mixture model}
\label{sec:tvc_gmm}

To further address residual multimodality, we increase the flexibility of $\tilde{\mathbf{Y}}$ by replacing $\tilde{\mathbf{Y}}^{t, f}$ with a mixture model as similarly proposed for low-dimensional acoustic features in~\cite{zenDeepMixtureDensity2014}:
\begin{align}
    \tilde{\mathbf{Y}}^{t, f} \sim \sum_{k=1}^K \alpha^{t,f}_k\mathcal{N}(\mathbf{\mu}^{t,f}_k, \mathbf{\Sigma}^{t,f}_k) \,,
\end{align}
where $K$ denotes the number of components and $\alpha_k^{t,f}$ denotes the mixing coefficients. 
In practice, we adapt the last network layer to predict the parameters $\alpha_k^{t,f}$, $\mathbf{\mu}^{t,f}_k$ and $\mathbf{\Sigma}^{t,f}_k$ and minimise negative log-likelihood (NLL) loss using the ground-truth spectrogram and two time/frequency shifted copies as targets (see Figure \ref{fig:tvc_gmm} for an illustration).
This is what we call Trivariate-Chain Gaussian Mixture Modelling (TVC-GMM).

\subsection{Sampling from TVC-GMM}
\label{sec:sampling}

We propose naive sampling and conditional sampling as two ways to sample a mel-spectrogram from TVC-GMM.
Naive sampling draws $\langle y^{t,f},y^{t+1,f},y^{t,f+1}\rangle$ from $\tilde{\mathbf{Y}}^{t,f}$ for each time/frequency bin in parallel.
As a result, we have multiple values at the overlap of the trivariate-chains, which we simply average to smooth sampling noise.
However, over-sharpness from noise can introduce new artefacts -- although vocoders tolerate it better than over-smoothness (see Section \ref{sec:vocoder_degradation}).
Thus, to further increase consistency and reduce sampling noise, conditional sampling uses an iterative algorithm which conditions each $\tilde{\mathbf{Y}}^{t, f}$ on the values drawn in previous bins. 
Instead of sampling $\langle y^{t,f}, y^{t+1,f}, y^{t,f+1}\rangle$ from $\tilde{\mathbf{Y}}^{t, f}$, we fix the known value of $y^{t,f}$ (predicted as $y^{t+1,f}$ in the previous bin) and obtain a bivariate slice of $\tilde{\mathbf{Y}}^{t, f}$ from which we only sample $\langle y^{t+1,f}, y^{t,f+1}\rangle$ consistent with the previous timestep.
The overlapping values in frequency direction are still averaged and we leave investigation into more sophisticated sampling approaches to future work.

\section{Experiments}

We conduct experiments on several datasets, vocoders and an adapted FastSpeech 2 model to show (1) over-smoothness causes audio degradation, (2) smoothness is exaggerated on expressive speech datasets and (3) TVC-GMM is effective in reducing smoothness and improving perceptual audio quality.

\begin{figure*}[ht]
    \centering
    \begin{subfigure}[b]{0.33\textwidth}
         \centering
         \includegraphics[width=1.1\textwidth]{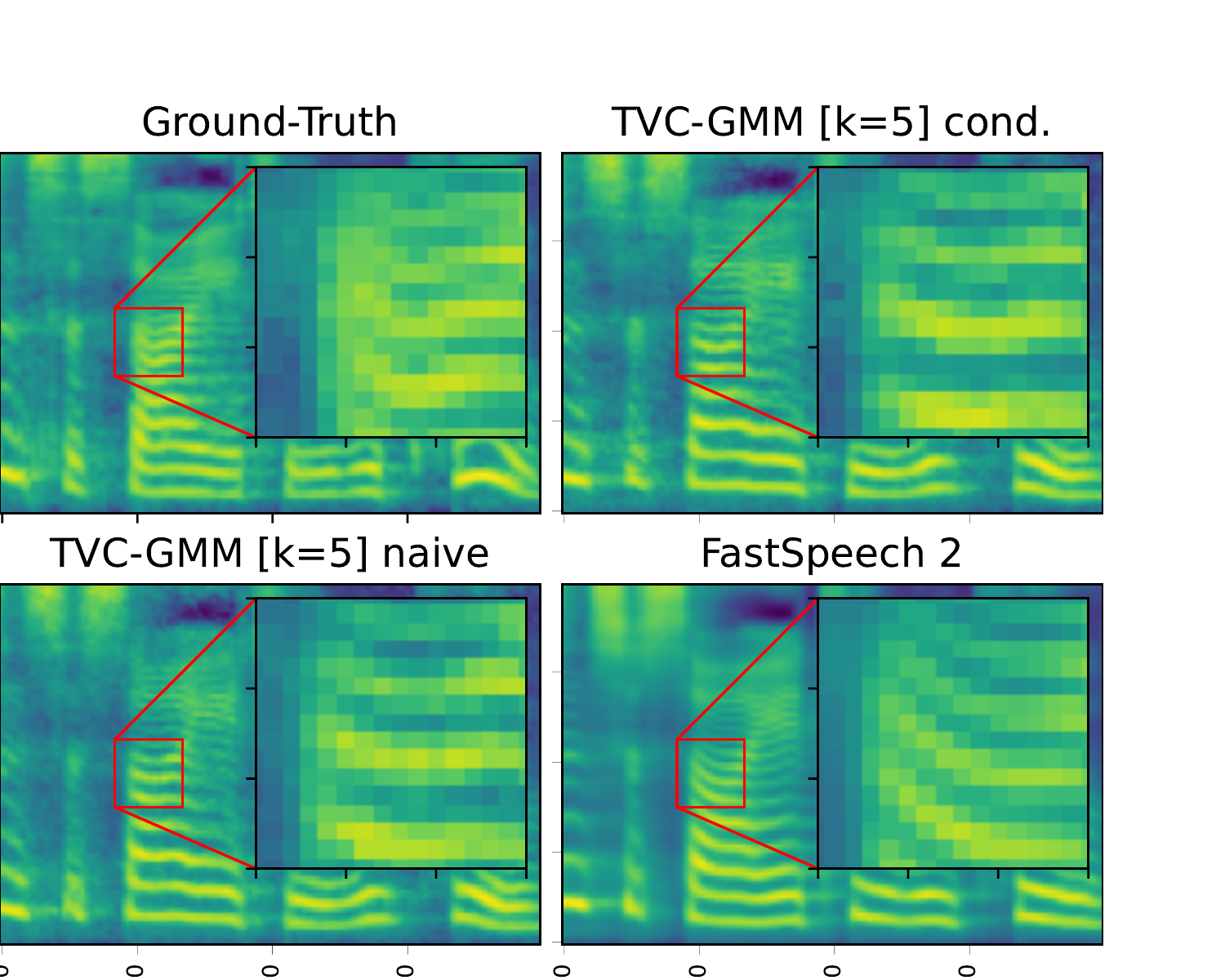}
         \caption{LJSpeech}
         \label{fig:ljs_mel}
     \end{subfigure}
     \hfill
     \begin{subfigure}[b]{0.33\textwidth}
         \centering
         \includegraphics[width=1.1\textwidth]{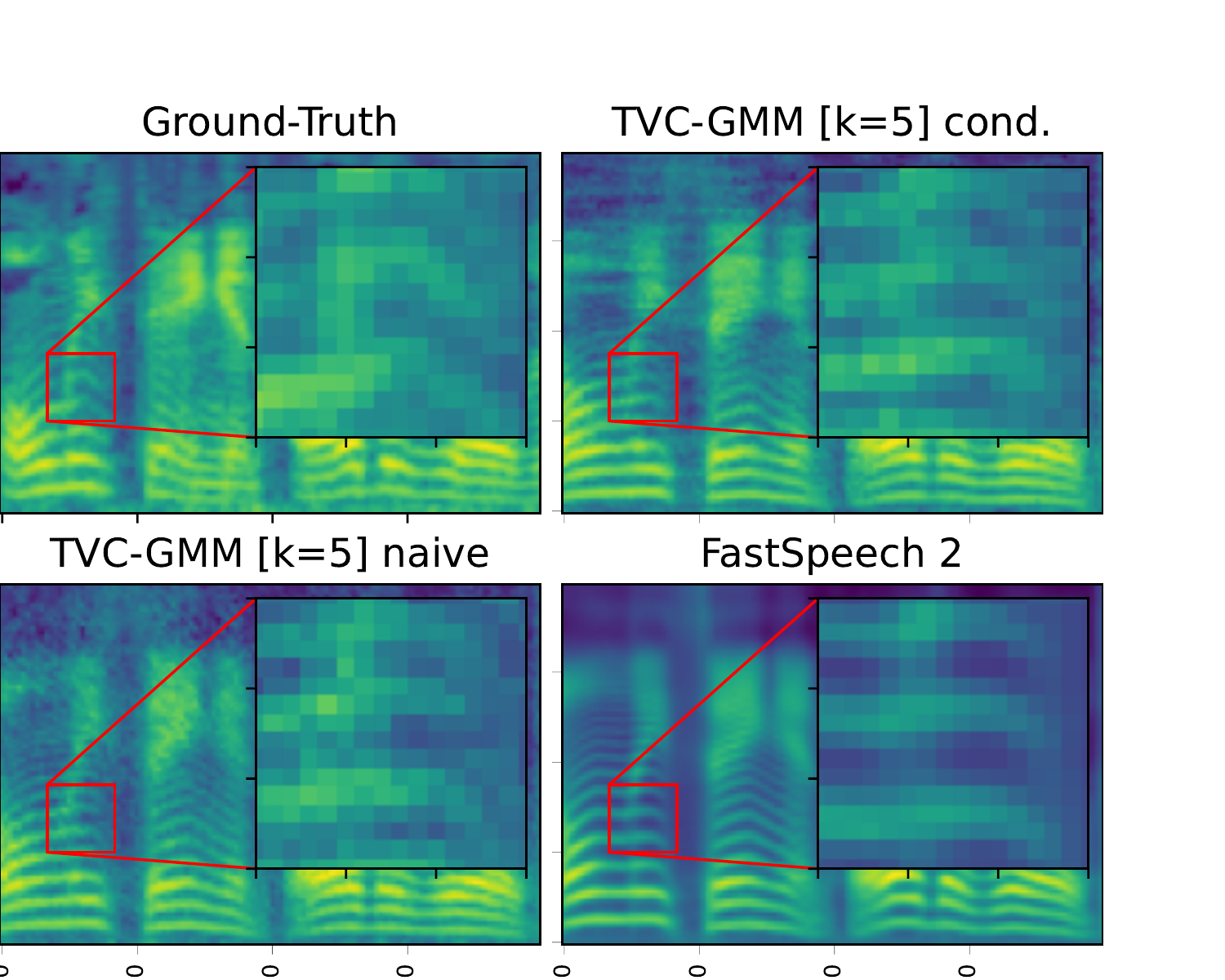}
         \caption{LibriTTS}
         \label{fig:ltts_mel}
     \end{subfigure}
     \hfill
     \begin{subfigure}[b]{0.33\textwidth}
         \centering
         \includegraphics[width=1.1\textwidth]{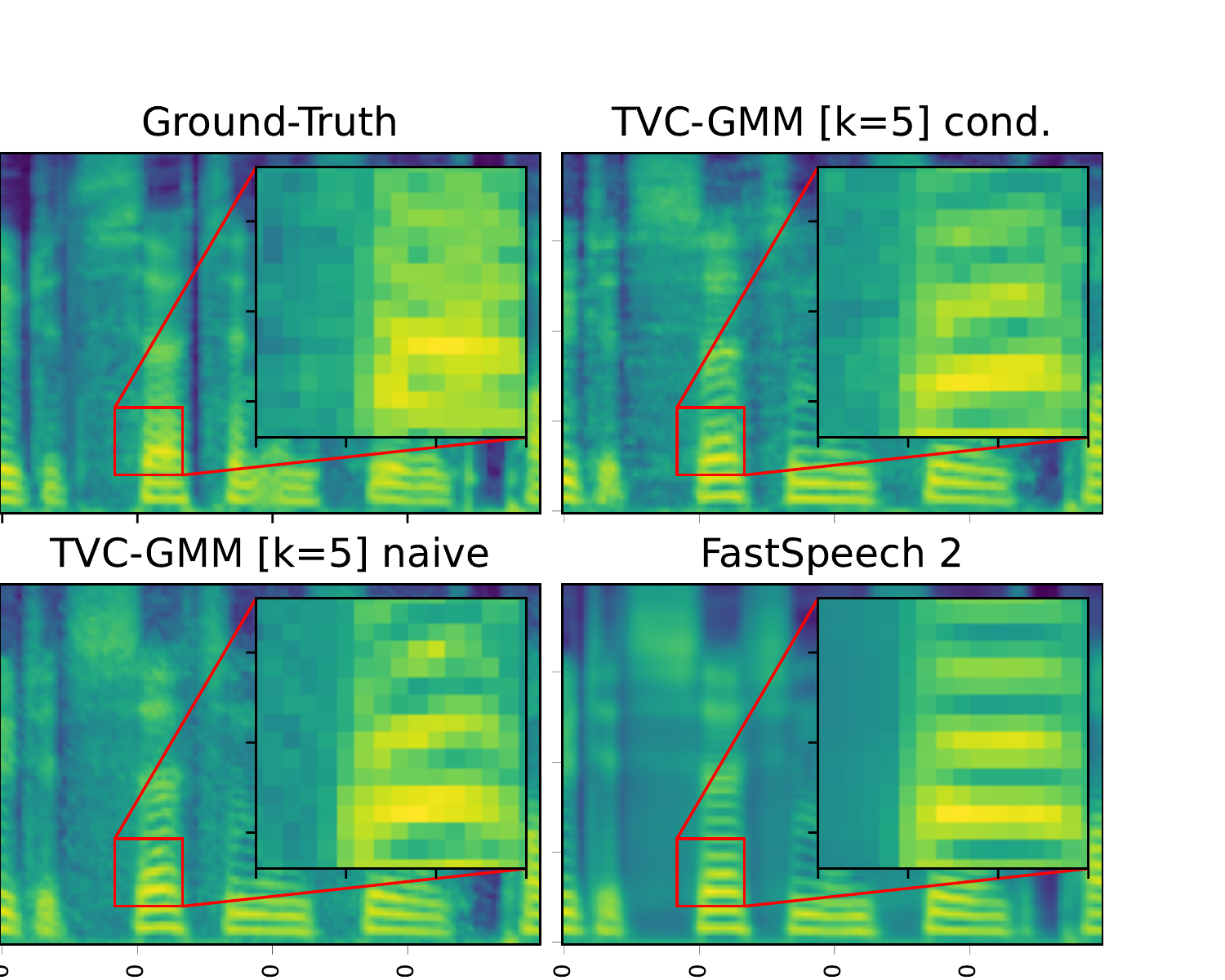}
         \caption{VCTK}
         \label{fig:vctk_mel}
     \end{subfigure}
  \caption{Aligned synthesized mel-spectrogram samples for all datasets and models. FastSpeech 2 models are visibly oversmooth, while TVC-GMM models are closer to ground-truth. Conditional sampling reduces the sampling noise introduced by naive sampling.}
  \label{fig:spectrograms}
\end{figure*}

\subsection{Datasets \& Model Details}

\begin{figure}[ht]
    \centering
    \includegraphics[width=\columnwidth]{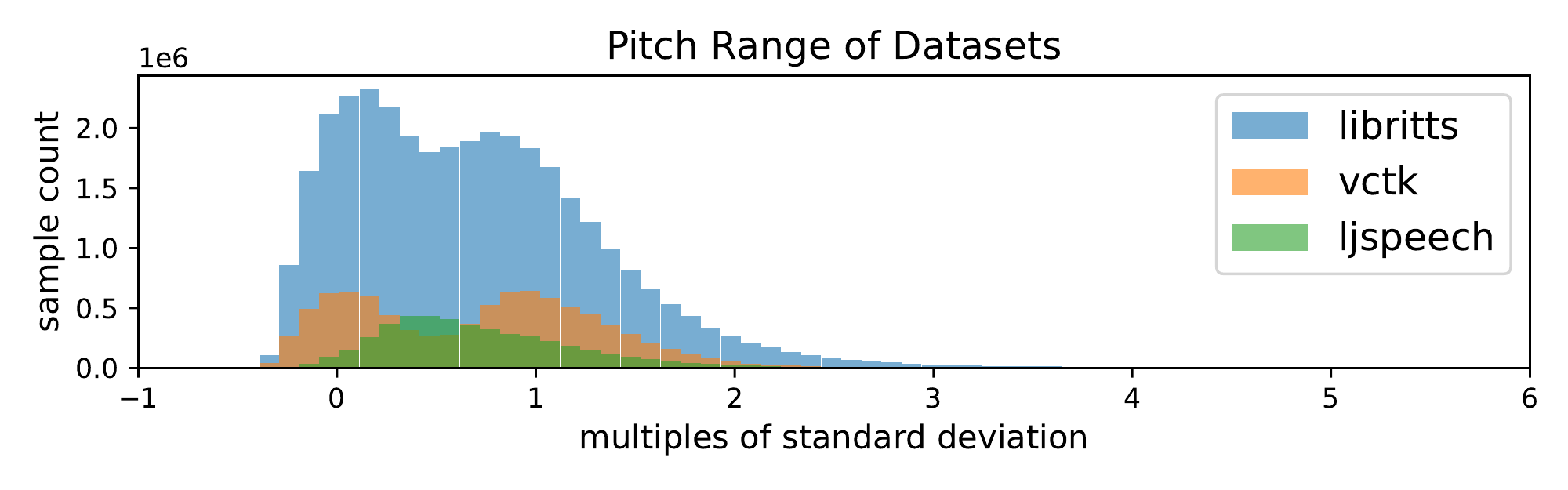}
    \caption{Pitch range of datasets. LibriTTS and VCTK are more diverse than LJSpeech.}
    \label{fig:dataset_pitch_range}
\end{figure}

We report results for three expressive English datasets.
We asses expressiveness by pitch range since FastSpeech 2 spreads its bins evenly (Figure \ref{fig:dataset_pitch_range}).
(1) LJSpeech~\cite{itoLJSpeechDataset2017} is the least expressive and contains 13k (24h) quite monotone samples from a single female speaker.
It is the standard evaluation set for many TTS models.
(2) VCTK~\cite{yamagishiCSTRVCTKCorpus2019} contains 88k (44h, ~20min/speaker) samples from 109 English speakers with various accents, deliberately selected for contextual and phonetic coverage.
(3) LibriTTS~\cite{zenLibriTTSCorpusDerived2019} is the standard multi-speaker dataset for TTS models.
We use the train-clean-360 split covering 115k (192h, ~15min/speaker) samples from 904 speakers and the largest expressive range in our comparison.
We set aside a test set of 512 samples per dataset for evaluation, spoken by seen speakers. 

We adapt the FastSpeech 2 implementation\footnote{we adapt \href{}{https://github.com/ming024/FastSpeech2/tree/d4e79e}} of~\cite{chienInvestigatingIncorporatingPretrained2021} as follows:
We reduce encoder and decoder stacks to 4 layers with 2 attention heads and remove the postnet as it is not in the original paper and preliminary experiments did not show significant impact. 
We also rearrange the variance predictors to run in parallel to better disentangle pitch and energy conditioning.
To further improve controllability, we feed the mean external conditioning signal into the predictor during training to encourage it to focus on the variation around this \enquote{prosody baseline}.
These changes do not affect the mechanism of variance conditioning, as we still learn a fixed set of pitch and energy embeddings.
However, during inference we can now control pitch and energy more targeted by setting the prosody baseline instead of scaling the predictors outputs.
As vocoder we use the HiFiGAN V1 universal checkpoint and the LJSpeech checkpoint for LJSpeech experiments.
We train all models 40k steps on a single Nvidia GeForce RTX
2080ti using NLL loss as described in Section \ref{sec:tvc_gmm}. 
FastSpeech 2 has 29.12M parameters, TVC-GMM [k=1] 29.43M (+1\%) and TVC-GMM [k=5] 30.25M (+4\%).
Average training time for TVC-GMM was $2.51\pm 0.97$ hours and for FastSpeech 2 $2.03\pm 0.87$ hours. 
TVC-GMM [k=5] is only 0.8\% slower in inference than FastSpeech 2, mainly due to the enlarged last layer and sampling.

\subsection{Insufficient modelling degrades reconstruction quality}
\label{sec:vocoder_degradation}

Modelling conditional averages in multi-valued mappings leads to smooth spectrogram predictions that also visually lie somewhere between two possible realisations (Figure \ref{fig:spectrograms}).
We use a filter kernel to artificially smooth (gaussian, $\sigma=1.0$) and sharpen (laplacian, $s=1$) the ground-truth spectrograms and calculate the change in perceptual speech quality~\cite{WidebandExtensionRecommendation2007} (PESQ) and perceptual audio distance~\cite{manochaCDPAMContrastiveLearning2021} (CDPAM) relative to the ground-truth audio (Table \ref{tab:vocoder_synth}).
Following~\cite{renRevisitingOverSmoothnessText2022}, we measure smoothness by the variance of the Laplacian-filtered mel-spectrogram $\text{Var}_L$~\cite{pech-pachecoDiatomAutofocusingBrightfield2000}.
A lower $\text{Var}_L$ indicates more smoothness.
We observe characteristic \enquote{metallic} (over-smooth) and \enquote{bubbling} (over-sharp) artefacts in the audio reconstruction across popular vocoder models\footnote{audio demo at \href{}{https://sony.github.io/ai-research-code/tvc-gmm}}.
Generally, over-sharpness is better tolerated.
As a naive way to reduce artefacts, we also experiment with fine-tuning HifiGAN (+finetuned) on smoothed LJSpeech to better tolerate smoothed spectrograms.
We find it increases smoothness tolerance on all datasets, however, it also decreases performance on GT spectrograms and requires additional effort for every practical application.

\begin{table*}[th!]
  \caption{Mel-spectrogram smoothness measured by variance of Laplacian $\text{Var}_L$~\cite{pech-pachecoDiatomAutofocusingBrightfield2000} causes reduction in perceptual speech quality (PESQ) and perceptual audio distance (CDPAM) relative to the ground-truth audio. Calculated over test sets (512 samples/dataset). Over-smoothness/-sharpness both induce audio artefacts across vocoder architectures, but over-sharpness is tolerated better.}
  \label{tab:vocoder_synth}
  \centering
  \resizebox{\textwidth}{!}{%
  \begin{tabular}{clr|rrrrrrr|rrrrrrr}
    \toprule
        && & \multicolumn{7}{c|}{\textbf{Change in PESQ\textsuperscript{*} $\uparrow$}} & \multicolumn{7}{c}{\textbf{CDPAM $\times 10^2$ $\downarrow$}}\\
    \midrule
    &&$\textbf{Var}_L$&
    \rotatebox{75}{\small\textbf{Griffin-Lim~\cite{perraudinFastGriffinLimAlgorithm2013}}}&
    \rotatebox{75}{\small\textbf{HiFiGAN~\cite{kongHiFiGANGenerativeAdversarial2020}}}&
    \rotatebox{75}{\small\textbf{+finetuned}}&
    \rotatebox{75}{\small\textbf{MelGAN~\cite{kumarMelGANGenerativeAdversarial2019}}}&
    \rotatebox{75}{\small\textbf{CARGAN~\cite{morrisonChunkedAutoregressiveGAN2022}}}&
    \rotatebox{75}{\small\textbf{WaveGlow~\cite{prengerWaveglowFlowbasedGenerative2019}}}&
    \rotatebox{75}{\small\textbf{WaveRNN~\cite{kalchbrennerEfficientNeuralAudio2018}}}&
    \rotatebox{75}{\small\textbf{Griffin-Lim~\cite{perraudinFastGriffinLimAlgorithm2013}}}&
    \rotatebox{75}{\small\textbf{HiFiGAN~\cite{kongHiFiGANGenerativeAdversarial2020}}}&
    \rotatebox{75}{\small\textbf{+finetuned}}&
    \rotatebox{75}{\small\textbf{MelGAN~\cite{kumarMelGANGenerativeAdversarial2019}}}&
    \rotatebox{75}{\small\textbf{CARGAN~\cite{morrisonChunkedAutoregressiveGAN2022}}}&
    \rotatebox{75}{\small\textbf{WaveGlow~\cite{prengerWaveglowFlowbasedGenerative2019}}}&
    \rotatebox{75}{\small\textbf{WaveRNN~\cite{kalchbrennerEfficientNeuralAudio2018}}}\\
    \midrule
    \multirow{3}{0.6em}{\rotatebox[origin=c]{90}{\scriptsize\textbf{LJSpeech}}} 
    & GT Spec.      & 0.37 & -1.40 & -1.48 & -1.66 & -2.63 & -1.73 & -1.09 & -1.56 & 26.1 & 6.0 & 10.6 & 9.3 & 19.6 & 5.5 & 5.7 \\
    & + Sharpen     & 0.66 & -1.78 & -1.86 & -2.73 & -2.77 & -2.15 & -1.40 & -1.89 & 24.0 & 8.2 & 19.4 & 11.6 & 17.1 & 6.5 & 6.9 \\
    & + Smooth      & 0.07 & -2.57 & -2.55 & -1.53 & -3.07 & -2.77 & -2.28 & -2.55 & 30.1 & 10.2 & 5.2 & 15.1 & 23.7 & 10.8 & 8.7 \\
    \midrule
    \multirow{3}{0.6em}{\rotatebox[origin=c]{90}{\scriptsize\textbf{VCTK}}} 
    & GT Spec.      & 0.36 & -1.37 & -1.64 & -1.86 & -2.49 & -1.28 & -1.46 & -2.23 & 27.3 & 13.8 & 19.3 & 14.7 & 8.4 & 20.1 & 20.8 \\
    & + Sharpen     & 0.77 & -1.76 & -2.02 & -2.6  & -2.66 & -1.82 & -1.68 & -2.54 & 26.6 & 16.2 & 25.6 & 14.4 & 9.7 & 18.2 & 18.5 \\
    & + Smooth      & 0.06 & -2.52 & -2.63 & -2.01 & -3.04 & -2.84 & -2.33 & -2.72 & 27.9 & 20.1 & 15.5 & 15.9 & 19.2 & 20.7 & 20.5 \\
    \midrule
    \multirow{3}{0.6em}{\rotatebox[origin=c]{90}{\scriptsize\textbf{LibriTTS}}} 
    & GT Spec.      & 0.41 & -1.51 & -1.69 & -2.03 & -2.66 & -1.67 & -1.63 & -2.24 & 27.8 & 12.1 & 17.5 & 10.2 & 13.3 & 17.7 & 17.7 \\
    & + Sharpen     & 0.66 & -1.79 & -2.06 & -2.83 & -2.82 & -2.11 & -1.82 & -2.49 & 27.4 & 14.3 & 24.9 & 11.8 & 13.3 & 16.3 & 16.6 \\
    & + Smooth      & 0.08 & -2.68 & -2.71 & -2.13 & -3.16 & -2.98 & -2.59 & -2.91 & 30.3 & 18.2 & 14.4 & 15.7 & 20.1 & 19.6 & 18.3 \\
    \bottomrule
    \multicolumn{17}{l}{\footnotesize * PESQ wide-band (16kHz), mapped to MOS-LQO (1.02 - 4.64 scale) following ITU-T P.862.2}\\
  \end{tabular}%
  }
\end{table*}

\subsection{Smoothness is exaggerated for expressive speech}
\label{sec:expressive_degradation}

Intuitively, the problem of over-smoothness can be exaggerated for expressive datasets and under prosody control when distributions are under-conditioned and more or further apart modes are averaged.
Indeed, in Table \ref{tab:tts_synth} we report lower $\text{Var}_L$ for LibriTTS samples generated by FastSpeech 2 than for LJSpeech samples and a larger drop in CDPAM relative to GT audio.
As Figure \ref{fig:dataset_pitch_range} demonstrates, LJSpeech has a comparably small pitch range and models trained on more expressive datasets such as VCTK and LibriTTS have to cover a larger diversity with the same number of conditioning bins.

\subsection{TVC-GMM improves perceptual audio quality}

To evaluate the effectiveness of TVC-GMM, we calculate the smoothness of synthesized mel-spectrograms and evaluate the reconstructed audio in objective and subjective studies.
As acoustic models we compare FastSpeech 2 and TVC-GMM models with 1 or 5 modes and naive or conditional sampling.
For objective evaluation only (Table \ref{tab:tts_synth}), we fix the phoneme duration prediction to align with the ground-truth audio and  report the perceptual audio distance CDPAM~\cite{manochaCDPAMContrastiveLearning2021}.
This metric is explicitly designed for TTS evaluation and shown to pick up on audio artefacts missed by other metrics, but perceived by humans.
For subjective evaluation (Table \ref{tab:tts_mos}), we randomly select 5 test samples per dataset, synthesise FastSpeech 2 and TVC-GMM audio and ask 19 proficient english speakers to rate the overall quality in terms of acoustic quality and prosodic diversity on a Likert-Scale of 1-5.
We let them compare between the models and include the ground-truth vocoder reconstruction as upper anchor as this is the best an acoustic model can achieve.

We find modelling with TVC-GMM reduces the smoothness of sampled spectrograms and in turn reduces the acoustic artefacts in the vocoder reconstruction.
This is evident in the first block of Table \ref{tab:tts_synth}, where a higher $\text{Var}_L$ indicates that spectrograms generated by TVC-GMM models are less smooth.
However, naive sampling from TVC-GMM yields over-sharp spectrograms with a $\text{Var}_L$ far above ground-truth, which we attribute to sampling noise.
Consequently, we use conditional sampling to reduce noise and improve perceptual audio similarity in particular on the expressive LibriTTS dataset (Table \ref{tab:tts_synth}, last column).
Despite only naive sampling being available for the MOS study, TVC-GMM outperforms in subjective evaluation as well.
Note, however, that a significant gap to the GT remains, especially for LibriTTS.
We believe this gap is largely due to the duration and variance prediction that control prosody.

\begin{table}[t]
  \caption{$\text{Var}_L$ and Perceptual Audio Distance (CDPAM) between ground-truth audio and HiFiGAN reconstruction from synthesis with TVC-GMM models and FastSpeech 2. Our TVC-GMM with both naive and conditional sampling outperforms. Bold values indicate $\text{Var}_L$ closest to GT and lowest CDPAM.}
  \label{tab:tts_synth}
  \centering
  \resizebox{\columnwidth}{!}{%
  \begin{tabular}{lrrr|rrr}
    \toprule
        &\multicolumn{3}{c}{$\textbf{Var}_L$}&\multicolumn{3}{c}{\textbf{CDPAM $\times 10^2$ $\downarrow$}}\\
    \midrule
        & \textbf{\footnotesize LJS}&\textbf{\footnotesize VCTK}&\textbf{\footnotesize LTTS}&\textbf{\footnotesize LJS}&\textbf{\footnotesize VCTK}&\textbf{\footnotesize LTTS}\\
    \midrule
    GT (HiFiGAN)        & 0.37 & 0.36 & 0.41 & 6.0 & 13.8 & 12.1 \\
    \midrule
    FastSpeech 2        & 0.23 & \textbf{0.25} & 0.21 & 11.0 & 19.7 & 22.1 \\
    TVC-GMM [k=1]       & 0.45 & 0.58 & 0.64 & 10.4 & 19.6 & 21.0 \\
    + cond. sampling    & \textbf{0.43} & 0.52 & 0.56 & 10.4 & \textbf{18.5} & 18.7 \\
    TVC-GMM [k=5]       & 0.44 & 0.60 & 0.63 & 10.2 & 19.5 & 20.1 \\
    + cond. sampling    & \textbf{0.43} & 0.56 & \textbf{0.55} & \textbf{10.1} & \textbf{18.5} & \textbf{18.5} \\
    \bottomrule
  \end{tabular}%
  }
\end{table}

\begin{table}[t]
  \caption{In subjective evaluation TVC-GMM (naive sampling) outperforms FastSpeech 2 in Mean Opinion Score (MOS).}
  \label{tab:tts_mos}
  \centering
  \begin{tabular}{l|rrr}
    \toprule
        &\multicolumn{3}{c}{\textbf{MOS $\uparrow$}}\\
    \midrule
        &\textbf{LJSpeech}&\textbf{VCTK}&\textbf{LibriTTS}\\
    \midrule
    GT (HiFiGAN)        & 3.87 ±0.02 & 3.90 ±0.03 & 3.80 ±0.05 \\
    \midrule
    FastSpeech 2        & 3.54 ±0.04 & 3.43 ±0.05 & 3.23 ±0.04 \\
    TVC-GMM [k=1]       & 3.69 ±0.03 & 3.61 ±0.04 & 3.39 ±0.04 \\
    TVC-GMM [k=5]       & \textbf{3.72 ±0.03} & \textbf{3.64 ±0.04} & \textbf{3.41 ±0.04} \\
    \bottomrule
  \end{tabular}%
\end{table}

\section{Conclusion \& Future Work}

Despite external pitch and energy conditioning, we observe audio artefacts caused by over-smooth mel-spectrogram predictions in FastSpeech 2, in particular in expressive speech datasets.
We conclude that this is due to insufficient modelling of the inherently multimodal natural speech distribution.
To model the residual multimodality after conditioning, we propose TVC-GMM which predicts trivariate-chain gaussian distributions in the mel-spectrograms and improves perceptual audio quality under objective and subjective evaluation. 
We target research and practical applications with a compromise between training effort and quality as TVC-GMM is naturally limited in the number of modes that can be modelled. 
We also acknowledge that TVC-GMM introduces a new class of sharpness artefacts, but attribute them to sampling noise, which unlike over-smoothness is not a model limitation.
We thus plan to investigate more sophisticated sampling strategies in future work. 

\clearpage
\bibliographystyle{IEEEtran}
\bibliography{bibliography}

\end{document}